# Cognitive Framework for Blended Mathematical Sensemaking in Science

Leonora Kaldaras[1], Carl Wieman[2]

[1]University of Colorado Boulder, Leonora.Kaldaras@colorado.edu

[2]Stanford Graduate School of Education, cwieman@stanford.edu

**Background**

Blended mathematical sensemaking in science ("Math-Sci sensemaking") involves deep conceptual understanding of quantitative relationships describing scientific phenomena and has been studied in various disciplines. However, no unified characterizations of Math-Sci sensemaking exists.

**Methods**

We developed a theoretical cognitive model for blended Math-Sci sensemaking grounded in prior work. The model contains three broad levels representing increasingly sophisticated ways of engaging in Math-Sci sensemaking: 1) deriving qualitative relationships among relevant variables describing a phenomenon ("qualitative level"), 2) deriving mathematical relationships among these variables ("quantitative level"), 3) explain how the mathematical operations used in the formula relate to the phenomenon ("conceptual level"). Each level contains three sublevels. We used PhET simulations to design dynamic assessment scenarios in various disciplines to test the model. We used these assessments to interview undergraduate students with a wide range of math skills.

**Findings**

Interview analysis provided validity evidence for the cognitive model. It also revealed that students tend to perform at the same level across different disciplinary contexts, suggesting that blended Math-Sci sensemaking is a distinct cognitive construct, independent of specific disciplinary context.

**Contribution**

This paper presents a first-ever published validated cognitive model for blended Math-Sci sensemaking which can guide instruction, curriculum, and assessment development.

## Introduction

Blended mathematical sensemaking in science ("Math-Sci sensemaking") is a special type of sensemaking that involves developing deep conceptual understanding of quantitative relationships and scientific meaning of equations describing a specific phenomenon (Zhao et al., 2021; Kuo et al., 2013). Blended Math-Sci sensemaking is an important component of expert understanding of science and expert mental models, and it is therefore a prerequisite for students building deep science understanding. While various aspects of the Math-Sci sensemaking have been described for specific disciplines (Bing & Redish, 2007; Tuminaro & Redish, 2007; Ralph & Lewis, 2018; Schuchardt, 2016; Lythcott, 1990; Schuchardt & Schunn, 2016; Hunter et al., 2021), there has been little work on formulating and testing a theory of mathematical sensemaking as a cognitive concept that applies across different scientific fields. This paper offers initial evidence that a unified blended Math-Sci framework is possible. Having a general framework for discussing, diagnosing, and supporting the development of Math-Sci sensemaking across disciplines will help improve instruction and assessment principles. This in turn can lead to better learning outcomes and development of deeper science understanding.

To design the unified framework for blended Math-Sci sensemaking we first conducted a literature review in various science fields focused on the first research question of the study: (RQ 1): *How can one characterize the different ways of engaging in blended Math-Sci sensemaking?*

We then developed a theoretical cognitive model ("framework") for blended Math-Sci sensemaking. The framework outlines qualitatively different levels reflecting increasingly sophisticated ways of engaging in blended Math-Sci sensemaking. Then, we investigated whether this theoretical framework indeed represents the various ways in which students engage in such sensemaking. This is the second research question of the study (RQ2): *To what degree does the validity evidence support the theoretical framework for blended Math-Sci sensemaking?*

To answer RQ2, we probe the levels of the framework by leveraging the capabilities of PhET simulations. Specifically, one of the key features of the sensemaking process is its dynamic nature focused on continuously revising an explanation based on new evidence to figure something out (Oden, Russ, 2019). The dynamic nature of PhET simulations provides a unique and suitable environment for assessing Math-Sci sensemaking skills. They offer a dynamic and interactive assessment environment that allows for accumulation of new evidence and feedback associated with changing parameters of the system in question. This supports revisions of explanations by calling on blended understanding of the scientific concepts and the underlying mathematical relationships.

In the context of blended Math-Sci sensemaking, the relevant mathematical equations represent processes described by specific variables. Simulations, in turn, represent a physical behavior with certain variables that control that behavior. The simulation allows learners to explore how the behavior depends on different variables, both qualitatively and quantitatively, therefore providing a meaningful context for engaging in blended Math-Sci sensemaking. Simulations provide a simplified (but not too simplified) system for exploring the mathematical complexity of the phenomenon described in the simulation. These features of simulations were the reason for choosing PhET simulations as the assessment context for testing our theoretical cognitive model for the blended Math-Sci sensemaking.

We designed an interview protocol aimed at probing the levels of the theoretical blended Math-Sci sensemaking framework in the context of PhET simulations spanning Physics, Chemistry, and energy conversion disciplinary contexts. The range of scientific contexts was chosen to explore the extent to which the Math-Sci sensemaking varied with context. We collected and analyzed interviews with 25 undergraduate science and non-science majors with a wide range of math skills to test the validity of the theoretical framework. The interview analysis provided evidence of the validity of the proposed theoretical framework.

**Literature Review**

      Blended sensemaking refers to the process of combining separate cognitive resources to generate a new, blended understanding (Fauconnier & Turner, 1998). In the context of blended Math-Sci sensemaking the two cognitive resources are the Scientific and the Mathematical knowledge respectively. The blended sensemaking process in this context therefore refers to the process of using both Math and Science cognitive resources to make sense of phenomena as opposed to using only one of the cognitive resources (e.g. either Science or Math sensemaking). Ability to engage in blended Math-Sci sensemaking reflects higher level, expert-like understanding (Redish, 2017) and has been shown to help students in solving complex quantitative problems in science (Schuchardt & Schunn, 2016).

      In order to successfully support students in developing blended Math-Sci sensemaking, it is important to understand what proficiency looks like at different levels of sophistication with respect to blending those two resources. At present, there has not been a coherent framework developed and validated for characterizing proficiency in blended Math-Sci sensemaking. However, there has been considerable work published on characterizing different ways students can engage in Mathematical and Scientific sensemaking separately as well as characterizing blended sensemaking from different educational perspectives that are not specifically related to defining and characterizing proficiency at different levels of sophistication (see, for example, a very detailed review of relevant literature by Zhao & Schuchardt, 2021).

      One of the biggest advances in developing a coherent framework for characterizing different sensemaking opportunities in Science and Math has been pursued by Zhao and Schuchardt (2021). Specifically, Zhao and Schuchardt (2021) have presented a framework that captures sensemaking opportunities for mathematical equations in science grounded in the review of relevant literature. The framework presents the sensemaking opportunities along two separate dimensions: Science sensemaking and Mathematics sensemaking. The categories within the dimensions are ordered theoretically to represent increasingly sophisticated levels of sensemaking. The framework presented by Zhao and Schuchardt is theoretical and has not been validated in practice. While the framework presented by Zhao and Schuchardt can be used for characterizing both Math and Science sensemaking and identify opportunities for blended sensemaking during instruction, the framework doesn't offer explicit guidance for supporting *blended* Math-Sci sensemaking at different levels of sophistication. In this study we were specifically interested in developing and providing validity evidence for a framework for *blended* Math-Sci sensemaking that would describe *blended* sensemaking process at different levels of sophistication. Current work builds on the work of Zhao and Schuchardt and further developing the two separate cognitive dimensions of Science and Math into a unified cognitive dimensions of blended Math-Sci sensemaking and defining increasingly sophisticated proficiency categories for the blended sensemaking.

      Gifford and Finkelstein (2020) developed a cognitive framework for mathematical sensemaking in Physics which focuses on describing the process of sensemaking and relating it to basic cognition. The framework described by Gifford and Finkelstein doesn't focus on

defining what proficiency in blended Math-Sci sensemaking looks like at different levels of sophistication, which is the focus of the framework described in this study. Specifically, the current study focuses on defining and distinguishing different levels of proficiency in blended Math-Sci sensemaking for assessing and scaffolding instruction

**Theoretical Framework**

*Developing Cognitive Model for Defining Proficiency*

A cognitive model (also called a model of cognition) describes how students represent knowledge and develop proficiency in a domain (National Research Council [NRC], 2001). Proficiency refers to describing what mastery looks like in a given domain. The understanding of how proficiency develops is essential for designing effective instructional and assessment strategies. Cognition models allow for empirical testing and valid interpretation of assessment results, aligning curriculum, instruction, and assessment with the purpose of helping students achieve higher proficiency in a given concept (NRC, 2001).

Blended Math-Sci sensemaking is a cognitive construct that has been studied in various fields of science. While it is an important component of expert-like understanding and expert mental models, there has been limited work on defining and validating a cognitive model describing proficiency in this construct. An important first step was presented in a recent review paper by Zhao and Schuhard (2021). The authors synthesized relevant literature on mathematical and scientific sensemaking as distinct entities to generate theoretical categories that capture sensemaking opportunities for mathematical equations in science. They provided categories divided into two dimensions (see Figure 1): science sensemaking and mathematics sensemaking (Zhao & Schuhard, 2021). Their science sensemaking dimension includes four categories organized in the order of increasing sophistication of understanding: scientific label ("Sci Label"), scientific description ("Sci Description"), scientific pattern ("Sci Pattern") and scientific mechanism ("Sci Mechanism"). The math sensemaking dimension includes five categories in order of increasing sophistication: "Math-Procedure", "Math-Rule", "Math-Structure", "Math-Relation" and "Math-Concept". (Zhao, Schuhard, 2021). Their ordering of the levels is based on the cognitive complexity required for engaging in various types of sensemaking. For example, logically, engaging in "Sci-Mechanism" type of sensemaking requires first being able to identify specific properties and the corresponding variables relevant to characterizing a given phenomenon ("Sci-Description"). Once the variables have been identified, it is possible to engage in identifying specific patterns among the relevant variables ("Sci-Pattern"). Finally, once the relevant patterns have been identified, it is possible to engage in developing a causal mechanistic account of the phenomenon ("Sci Mechanism"). This leads to the increasing order of sophistication described by Zhao and Schuhard for science sensemaking dimensions. Similar logic applies to the mathematics sensemaking dimension. Zhao & Schuchardt note the need to empirically test these levels of sophistication for both dimensions. The current work extends the work of Zhao and Schuchardt (2021) and their categories of sophistication to develop and empirically validate a cognitive model that combines the two dimensions to achieve *blended* Math-Sci sensemaking.

*Developing Theoretical Cognitive Model for Blended Math-Sci Sensemaking*

We used a subset of the science and mathematics sensemaking categories described by Zhao and Schuhard (Z & S) and blended them together to design new categories that each combine a mathematics and a science dimension to reflect the blended nature of the cognitive model. The blending process is illustrated in Figure 1 and discussed in detail below. Since the focus of the cognitive framework is quantitative understanding of scientific phenomena, the math sensemaking dimension is given precedence in the development of the blended categories. This was based on our belief, which the data confirmed, of the central role that quantitative understanding plays in blended Math-Sci sensemaking construct proficiency.

In blending the dimensions, we chose not to use the lowest two categories of Z & S Mathematics sensemaking dimensions ("Math Procedure" and "Math Rule") and the lowest category of science sensemaking dimensions ("Sci Label"). The reason for not using those categories was that they represent basic sensemaking skills in each dimension and would not allow for any meaningful type of blended sensemaking. Specifically, "Math Label" and "Math Rule" categories refer to student knowledge of Math procedures and rules respectively. While these categories are important examples of mathematical sensemaking, and an essential prerequisite for engaging in blended Math-Sci sensemaking, these categories are very basic and do not allow for any meaningful type of blended sensemaking. Similarly, "Sci Label" refers to the student's ability to label variables and relate each variable in the equation to quantifiable aspects of the phenomenon. This category is largely definitional and doesn't allow for a meaningful blended sensemaking. Instead, we included it as an aspect of the "Math Structure-Sci Description" blended category below.

Each blended category was developed by combining the three math sensemaking dimension categories described by Z & S ("Structure", "Relation" and "Concept") with three Science sensemaking dimension categories ("Description", "Pattern", "Mechanism"). In other words, each of the three math sensemaking categories ("Structure", "Relation", "Concept") used in the study was subdivided into three Science sensemaking categories ("Description", "Pattern", "Mechanism"). The order of sophistication in the blended sensemaking followed that suggested by Z & S. As shown in Figure 1, the lowest math sensemaking category termed "Structure" was combined with each of the three science sensemaking categories to yield three blended Math-Sci sensemaking categories shown as the lowest ("qualitative") sublevel of the blended framework. Similar logic applied to blending each of the three science sensemaking categories with the two higher level math sensemaking categories including "Relation" and "Concept". The nine resulting categories are divided into three broad levels with respect to the math sensemaking dimension which we label as "qualitative", "quantitative", and "conceptual". These reflect different levels of proficiency in quantitatively describing phenomena. Each broad level contains three sub-categories reflecting the science sensemaking dimension as shown in Table 1. The resulting framework consists of new, categories that are adapted from the categories proposed by Z & S and follow similar ordering but reflect proficiency in *blended* Math-Sci sensemaking.

*Figure 1. Science and Mathematics categories of Zhao and Schuhard (2021) blended into new categories making up the blended Math-Sci framework.*

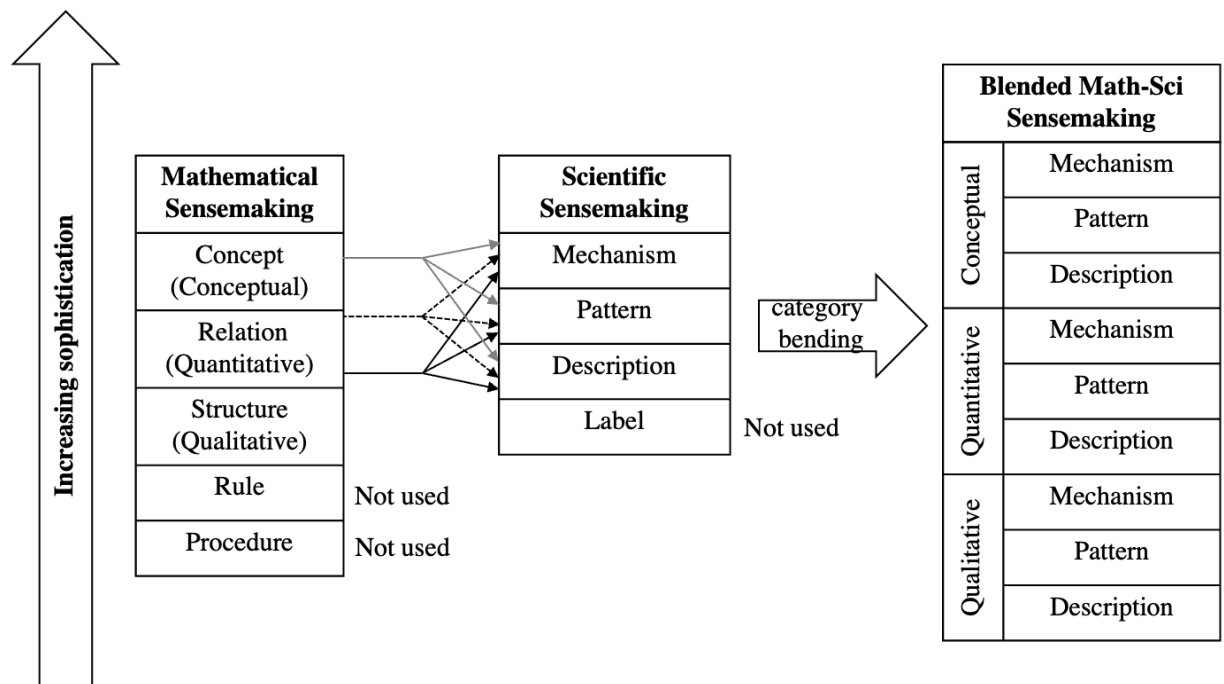

*Theoretical Cognitive Model for Blended Math-Sci Sensemaking*

    The detailed description of the blended categories is shown in Table 1. The lowest level, "qualitative", reflects the ability to identify *qualitative* aspects that are important for characterizing the phenomenon mathematically. The distinctive feature of this level is the qualitative nature of sensemaking. While students can engage in sensemaking of various aspects of the phenomenon in question, their sensemaking is limited to qualitative conclusions.

    The intermediate level, "quantitative", reflects the ability to develop a quantitative description of the phenomenon (e.g. develop a formula describing the phenomenon) and the sublevels mirror those at the qualitative level but reflect the ability to go beyond qualitative accounts. The distinctive feature of this level is the ability to identify quantitative relationships among the variables and translate these relationships into the appropriate mathematical operation. At this level students justify the mathematical operation of choice by recognizing that the observed quantitative patterns among the numerical values of the variables suggest that specific mathematical operation.

    Finally, the highest level, "conceptual", indicates a causal understanding of quantitative relationships. At this level students can identify additional unobservable variables needed to characterize the phenomenon mathematically and justify their choices. Students also justify the mathematical operations in the resulting equation by explicitly relating their quantitative observations to mathematical operations.

*Table 1. Theoretical Blended Math-Sci Sensemaking Framework*

| 1 Qualitative | Description | Students can use observations to identify which measurable quantities (variables) contribute to the phenomenon.<br>*Example: force and mass make a difference in the speed of a car.* |
|---|---|---|
| | Pattern | Students recognize patterns among the variables identified using observations and can explain *qualitatively* how the change in one variable affects other variables, and how these changes relate to the scientific phenomenon in question.<br>*Example: the smaller car speeds up more than the big car when the same force is exerted on both.* |
| | Mechanism | Students demonstrate *qualitative* understanding of the underlying causal scientific mechanism (cause-effect relationships) behind the phenomenon based on the observations but can't define the exact mathematical relationship.<br>*Example: it is easier to move lighter objects than heavy objects, so exerting the same force on a lighter car as on a heavy car will cause the lighter car to speed up faster.* |
| 2 Quantitative | Description | Students recognize that the variables identified using the observations provide measures of scientific characteristics and can explain *quantitatively* how the change in one variable affects other variables (but not recognizing the quantitative patterns yet), and how this change relates to the phenomenon in question. Students not yet able to express the phenomenon as an equation.<br>*Example: recognizing that as variable A changes by 1-unit, variable B changes by 2 units.* |
| | Pattern | Students *recognize quantitative patterns* among variables and explain *quantitatively (in terms of an equation or formula)* how the change in one parameter affects other parameters, and how these changes relate to the phenomenon in question. Students not yet able to relate the observed patterns to the operations in a mathematical equation and can't develop the exact mathematical relationship yet.<br>*Example: recognizing linear and inverse relationships* |
| | Mechanism | Students can explain *quantitatively* (express relationship as an equation) for how the change in one variable affects other variables based on the quantitative patterns derived from observations. Students include the relevant variables that are not obvious or directly observable. Students not yet able to explain conceptually why each variable should be in the equation beyond noting that the specific numerical values of variables and |
| | | observed quantities match with this equation. Students cannot explain how the mathematical operations used in the equation relate to the phenomenon, and why a certain mathematical operation was used. Students can provide causal account for the phenomenon. <br> _Example_: _In $F_{net}=m*a$, multiplication makes sense because when applied force on the mass of 50 kg increases from 10 to 20 N, acceleration increases by 2. That only makes sense for a multiplication operation._ |
| **3** <br> **C** <br> **o** <br> **n** <br> **c** <br> **e** <br> **p** <br> **t** <br> **u** <br> **a** <br> **l** | Description | Students can describe the observed phenomenon in terms of an equation, and they can explain why all variables or constants (including unobservable or not directly obvious ones) should be included in the equation. Students not yet able to explain how the mathematical operations used in the formula relate to the phenomenon. <br> _Example_: _In $F=m*a$, the F is always less than applied force by specific number, so there must be another variable subtracted from $F_{applied}$ to make the equation work. The variable involves the properties of the surface. So, the equation should be modified: $F_{applied}-(variable)=m*a$_ |
| | Pattern | Students can describe the observed phenomenon in terms of an equation, and they can explain why all variables or constants (including unobservable or not directly obvious ones) should be included in the equation. Students not yet able to provide a causal explanation of the equation structure. <br> _Example_: _In $F_{net}=m*a$, multiplication makes sense because as applied force on the same mass increases, acceleration increases linearly, which suggests multiplication._ |
| | Mechanism | Students can describe the observed phenomenon in terms of an equation, and they can explain why all variables or constants (including unobservable or not directly obvious ones) should be included in the equation. Students can fully explain how the mathematical operations used in the equation relate to the phenomenon in questions and therefore demonstrate _quantitative_ conceptual understanding. Here, the conceptual understanding is how mathematical relationships represent numerical dependencies. <br> _Example_: _since greater acceleration is caused by applying a larger net force to a given mass, this shows a positive linear relationship between a and $F_{net}$, which implies multiplication between m and a in the equation, or $F_{net}=m*a$._ |

*Validating the theoretical cognitive model for blended Math-Sci sensemaking*

Validating a cognitive model starts with developing a theoretical model reflecting what different levels of proficiency look like in a domain. This model was presented above. We empirically tested the model using assessment interview scenarios from three different subject domains (Physics, Chemistry and Energy Conversion). These interviews probed how well student thinking fit within the levels and sublevels of the framework shown in Figure 1 and Table 1. The student responses were the data used to test the validity of the model. This method is following the Standards for Educational and Psychological Testing as appropriate for test validity evidence (American Educational Research Association [AERA], 2018) and has been previously used to validate cognition models such as learning progressions (Briggs & Alonzo, 2012; Briggs et al., 2006; Mohan et al., 2009).

Response process-based validity is obtained by evaluating the correspondence between responses to assessments measuring the construct for a population of students and the and the various cognitive model levels. If there is sufficient evidence of correspondence between the variation in student responses and the theoretical model levels, one can conclude that a given cognitive model exhibits response process-based validity. We used a sample of students with a wide range of math and science proficiency to adequately test the model.

**Methods**

To test the theoretical cognitive model for blended Math-Sci sensemaking, we first selected suitable PhET simulations spanning three subject areas that would be suitable to provide assessment scenarios. Then we developed an interview protocol for each scenario that would probe each level and sublevel of the model. That was followed by interviewing undergraduate students spanning a range of majors and mathematical proficiency on all three scenarios, then coding their responses and comparing with the levels of the theoretical model. We discuss each step in more detail below.

*Choosing disciplinary contexts and PhET simulations*

Most studies on studying Mathematical Scientific and blended sensemaking have been conducted in the fields of Physics, Chemistry and Biology (Zhao & Schuchardt, 2021). Therefore, the theoretical model for blended sensemaking shown in Table 1 is largely built on the categories developed by Z & S that in turn are based on prior work in those fields. We initially planned to use those three disciplines as a context for the current study. The next step was to choose an appropriate phenomenon in each of those fields. The main criteria were: 1) the simplicity of the mathematical relationship describing the phenomenon; 2) the observational simplicity of the phenomenon; 3) the wide applicability of the scientific idea underlying the phenomenon. As described below, we could not find suitable assessment scenarios for biology that met all the other criteria, so we ended up choosing a different field for the third scenario.

In terms of mathematical simplicity, the criteria were phenomena that were described by a simple mathematical relationship (e.g. linear or inverse multiplicative relationships). This allowed a substantial fraction, though far from all, of the interviewees (and presumably our target population) to express the mathematical relationship based on their interactions with the PhET simulation describing the phenomenon. We also chose phenomena described by a mathematical relationship with only one type of mathematical operation to reduce the mathematical complexity. We also chose PhET simulations that model phenomena that most students are familiar with from everyday life and/or their coursework. Finally, we chose phenomena that are based on a widely applicable science ideas, so that the assessment might offer a useful learning opportunity to the student volunteers.

PhET simulations offer uniquely suitable environment for assessing blended Math-Sci sensemaking skills by providing a dynamic, interactive system governed by mathematical equations. PhET simulations provide an assessment environment that allows for continuous accumulation of evidence and feedback associated with changing parameters of the system in question, which is central to supporting the sensemaking process (Oden & Russ, 2019). Specifically, in the context of blended Math-Sci sensemaking, the relevant mathematical equations represent processes described by specific variables. PhET simulations features make it possible to represent a physical behavior with certain variables that control that behavior. The simulations allow learners to explore how the behavior depends on different variables, both qualitatively and quantitatively, therefore providing a meaningful context for engaging in blended Math-Sci sensemaking. PhET simulations therefore offer unique supports for revising explanations by calling on blended understanding of the scientific concepts and the underlying mathematical relationships.

Following these criteria, we chose a PhET simulation modeling acceleration on an object as a function of applied force (Newton's Second law) for Physics. The phenomenon is described by the formula $F_{net}$=ma, where "F" is a net force exerted on an object (calculated by subtracting applied force from the force of friction), "m" is mass of an object and "a" is the acceleration of the object. The formula involves a simple linear relationship and describes a familiar phenomenon. For Chemistry, we chose a PhET simulation modeling the relationship between concentration of a substance and the resulting absorbance at a given wavelength (Beer's law). The phenomenon is described by the formula A = c b  e, where "A" is the absorption at a given wavelength, "c" is the concentration of a substance, "b" is the width of the substance's container and "e" is a molar absorption coefficient constant reflecting an internal property of a substance.

We could not find any simple Biology phenomena that would meet the criteria discussed above, particularly the appropriate level of mathematical complexity. Most widely used formulas describing biological phenomena, such as population growth, involve complex relationships that cannot be easily derived by a typical student within the limited timeframe of an interview. Therefore, we decided not to pick a phenomenon in Biology.

This led us to select a more interdisciplinary phenomenon for our third scenario, the conversion of energy and the efficiency of the conversion across different systems. The phenomenon is described by an efficiency formula which could be represented in one of the two ways: a) *Fraction of the Energy Used=Useful Energy output/Energy input* or b) *Useful Energy Output=Energy Input- Energy Lost to useless forms*. This phenomenon was chosen because it represented a simple inverse mathematical relationship, which is different from Chemistry and Physics disciplinary contexts that were both described by linear relationship formulas. Further, the phenomenon involves familiar and important ideas and contexts that span different disciplines. Finally, the PhET simulation showing energy conversion allowed learners to investigate the energy efficiency of various systems. We label this context "Energy Conversion". We will further explain interview protocol development for these disciplinary contexts represented by the PhET simulations described here.

*Developing interview protocol*

We developed an interview protocol that was used for all three assessment scenarios (Physics, Chemistry, Energy Conversion). The interview questions probed the levels of the theoretical cognitive model for blended Math-Sci sensemaking shown in Table 1. The interview protocol is provided in the Appendix.

The interview questions focused on asking students to use the PhET simulation to explore the phenomenon and then characterize the behavior mathematically. Then, a set of questions probed the mastery of the lowest level of the framework by asking students to identify the relevant variables, note qualitative patterns among the variables, and qualitatively explain causal relationships between the variables.

Next, the student thinking at the intermediate ("quantitative") level was probed by asking students to determine the numerical values of the relevant variables and the quantitative patterns among the variables. This was followed by asking them to develop mathematical relationships (express a mathematical relationship or equation) among the variables and explain the justification for that quantitative relationship.

At this point, if students were struggling to provide a mathematical relationship based on their interaction and observations with the simulation, they were provided with data that was collected from the simulation (or relevant to the simulation, as the case with Energy Conversion). This data was presented to them in a table which reflected how numerical values of the relevant variables change with respect to each other. For example, for the Acceleration simulation, students were provided with three data tables: one showing how acceleration changes as different forces are applied to the same mass; a second showing how acceleration changes as the same force is applied to different masses; and a third table showing the acceleration for the combination of different objects with the same resulting mass (to demonstrate that it is the resulting mass that matters, and not the combination of objects). As students were studying the data provided to them, they were also allowed to go back and forth between the data and the simulation to see if that helps them figure out the mathematical equation.

If they were still unable to give a suitable equation, they were presented with a list of possible equations and asked to use the simulation and the data to see if they can figure out which of these equations properly described the phenomena. The list of possible formulas was purposefully made very long with similar combinations of variables to minimize the possibility of guessing the correct formula simply based on which variables it contained. The data tables with lists of possible formulas for all the assessment scenarios are provided in the Appendix.

Finally, those students who mastered levels 1 and 2, by giving a correct equation based on either the PhET simulation alone, or a combination of the data and the list of possible formulas provided to them were assessed as to their mastery of level 3, conceptual. The students were asked to justify the mathematical relationships among the variables in the equation, and explicitly relate the mathematical operations to the observations in the simulation. This probed whether students could accurately translate observation patterns to specific mathematical operations. Additionally, students were also asked to provide a causal explanation for the equation structure that they proposed, focusing on probing whether students understand cause-effect relationships reflected in the equation.

*Interviews*

Interviews were conducted via zoom using standard zoom recording features. Each interview lasted between 40-60 minutes during which students were given time to interact with the simulation and answer interview questions provided in the Appendix. Each student completed three interviews, each focusing on one of the three subject areas and PhET simulations, respectively.

*Participants*

Participants were first- and second-year undergraduate students recruited from a large public university and a private university in Western US. Participants were chosen from the list of volunteers to represent a sample of students with varying levels of Math and Science preparation. The participants were recruited by sending an email to the list of volunteers introducing the interview opportunity and asking volunteers to sign up. A total of 26 students were interviewed. One student was dropped from the interview analysis because they didn't finish all three interviews. The relevant information on the participants' level of preparation is shown in table 2. All participants were compensated for each interview with $20 gift cards.  The study was approved by University of Colorado IRB protocol # 13-0455.

*Interview Analysis*

Interview analysis was conducted using rubrics designed for each of the three PhET simulations. Each rubric was aligned to the framework shown in Table 1 and described in detail what student responses should contain at each sub-level for each specific simulation. An example of Beer's law rubric is shown in Table 3. The Newton's Law and Energy Conversion rubrics are provided in the Appendix. The rubrics were reviewed by disciplinary and educational experts to ensure alignment with the framework. All interviews were analyzed by the first author following the respective rubrics. Inter-rater reliability (IRR) was obtained by having a science education researcher unfamiliar with the project code two interviews in each discipline using the respective rubric. All six interviews were selected from different students. The researcher coded each interview and indicated evidence of each sub-level in student responses by a timestamp. For Beer's law and Energy Conversion simulations, the IRR was 100% on the first try. For Acceleration simulation, the IRR was 100% following the discussion. When discussing rating for each interview, the researchers compared final level assignment and evidence for all other levels of the sensemaking framework detected in the interview. The raters compared timestamps to ensure that the same information from student responses is taken as evidence for all level assignments.

*Table 2. Participants' level of preparation.*

| Student | Physics | SAT Math Score | Major |
|---------|---------|----------------|-------|
| 1 | algebra-based | Not available | Undecided |
| 2 | algebra-based | 540 | X-Ray Technician |
| 3 | calculus-based | 750 | Aerospace Engineer |
| 4 | calculus-based | 760 | Integrated Physiology |
| 5 | calculus-based | 720 | Mechanical Engineer |
| 6 | calculus-based | 680 | Mechanical Engineer |
| 7 | calculus-based | Not available | Mechanical Engineer |
| 8 | algebra-based | 740 | Mechanical Engineer |
| 9 | calculus-based | Not available | Mechanical Engineer |
| 10 | algebra-based | Not taken | Civil Engineer |
| 11 | calculus-based | 790 | Biology |
| 12 | Honors Physics | 560 | Undecided |
| 13 | calculus-based | 760 | Human Biology |
| 14 | calculus-based | 730 | Symbolic Systems |
| 15 | calculus-based | 760 | Computer Science/Astronomy |
| 16 | calculus-based | 690 | Environmental System Engineering |
| 17 | calculus-based | 670 | Environmental systems engineering |
| 18 | calculus-based | 660 | undecided/may be electrical engineering |
| 19 | calculus-based | 760 | Symbolic systems/pre-med |
| 20 | calculus-based | Not taken | undecided/plan to do Physics |
| 21 | algebra-based | 710 | Elementary Education |
| 22 | HS Physics | 490 | Elementary Education |
| 23 | algebra-based | 560 | Elementary Education |
| 24 | None | 480 | Elementary Education |
| 25 | None | 580 | Elementary Education |

*Table 3. Interview coding rubric for Beer's Law simulation*

| | | |
|---|---|---|
| **1** | Description | Students identify concentration and width of the cuvette as variables that affect absorbance and transmittance. |
| | Pattern | Student identifies that for specific wavelength the larger the concentration the larger the absorbance, and the smaller the transmittance. |
| | Mechanism | Students recognize that the concentration of substance is the main causal factor behind the changing absorbance and transmittance but can't define the exact mathematical relationship for Beer's law. |
| **2** | Description | Students quantitatively describe how the change in concentration and cuvette width affects absorbance and transmittance but don't recognize quantitative patterns yet. <br> *Example: when I use concentration X for substance A, the absorbance changes to Y.* |
| | Pattern | Students recognize that the relationship between concentration/cuvette length and absorbance is positive linear, and between concentration/cuvette length and transmittance is not linear (may say logarithmic or inverse). Students not yet able to relate the observed patterns to the operations in a mathematical equation and can't develop exact mathematical relationship for Beer's law yet. |
| | Mechanism | Students can explain *quantitatively* (express the relationship as an equation) for how the change in concentration and cuvette width affects absorbance. The formula derived: A=concentration*width of vial*molar absorption coefficient (MAC). Students can't fully explain why MAC should be included in the equation and can't justify multiplication operations beyond the fact that numerical values of the variables otherwise don't agree. Students recognize that the cause for changing absorbance is concentration of the substance. <br> <u>Note</u>: *MAC is an unobserved variable because it is not reflected in the PhET simulation and can only be inferred by noticing that absorbance at a given concentration and wavelength is different across various substances. MAC is provided to students in data tables.* |
| **3** | Description | Students can express the relationship as an equation for absorbance (A=concentration*width of vial*molar absorption coefficient (MAC)) and explain that MAC relates to specific properties of a given substance, and therefore should be included in the equation. Students can't explain why multiplication is their operation of choice beyond the fact that the numerical values of the variables otherwise don't agree. |

| | Pattern | Students can develop the relationship as an equation for absorbance and explain how the patterns among variables in the formula relate to observations. Specifically, students recognize that concentration and container width have a positive linear relationship to absorbance, which suggests multiplication operation. They also recognize that concentration and container width relate to absorbance through the factor of MAC, which also suggests multiplication operation. Students not yet able to provide a causal explanation of the equation structure. |
|---|---|---|
| | Mechanism | Students recognize that the cause for the change in absorbance is primarily the change in concentration (all other factors such as cuvette width and MAC being related to concentration) and can relate all the variables and operations in the equation to the observations of the phenomenon. |

**Results**

Below are sample responses for each sub-level of the framework for all three disciplinary scenarios. Table 4,5 and 6 show sample responses for the sub-levels of "description", "patterns" and "mechanism" respectively at each of the broad levels ("qualitative", "quantitative" and "conceptual"). We were able to identify evidence of blended sensemaking corresponding to every level and sub-level of the framework.

Table 4 shows sample responses for each of the three disciplinary scenarios for the "Description" sub level at each of the broad levels of the framework. As you can see, the "description" sub-level at the lowest qualitative level reflects student ability to identify the observable variables relevant for characterizing the phenomenon mathematically. For *Acceleration* simulation, the variables are acceleration, mass, and force. For *Beer's Law* simulation the variables are concentration, container width and absorbance. For *Energy Conversion* simulation the variables are energy input and energy output.

The "description" sub-level at the next level, "quantitative", reflects the student's ability to notice the values of the variables corresponding to a particular situation, but not noticing the patterns of behavior corresponding to changes in any of those variables. For *Acceleration* simulation, this involves noticing there are specific numerical values for acceleration, mass, and force. For *Beer's Law* simulation this means noticing the specific numerical values for absorbance, concentration, and container width in one case. For *Energy Conversion* simulation this involves noticing the numerical values for the variables of energy input and output. At this level students can connect their qualitative observations with the specific numerical values of the variables from the sim or data provided to them (like with *Beer's Law* and *Energy Conversion* examples), but they don't notice any quantitative relationships between the variables (compare to the "patterns" sub-level of quantitative level shown in Table 5 where students can recognize specific numerical patterns from the data or the simulation).

Finally, the "description" sub-level at the highest level, "conceptual", reflects the student's ability to develop the exact mathematical relationship representing the phenomenon in question, including identifying the variables that are not directly observed or those that are not directly obvious. Students at this level can justify their mathematical relationship by stating that it is supported by the patterns among the numerical values of the relevant variables.

For *Acceleration* simulation, this involves specifically recognizing that $F_{net}$ is calculated by subtracting the applied force from friction force and justifying the mathematical relationship ($F_{net}=ma$) by using numerical values of the variables to show that the formula works. Note that recognizing that the friction force should be accounted for when calculating the acceleration resulting from an applied force exerted on a mass is not immediately obvious. The PhET simulation provide an easier way to observe applied force and net force ($F_{applied}$-$F_{friction}$) as separate variables, therefore simplifying the cognitive step of having to recognize an additional variable ($F_{friction}$) that should be a part of the equation. Similarly, in real life it is challenging to differentiate between speed and acceleration, since those two variables are closely related, and not directly distinguishable when observing an object moving. For example, pushing a toy car with varying applied force one might think that speed should be part of the mathematical relationship since it is also affected by the applied force. PhET simulation also simplifies the cognitive step of recognizing acceleration as a different variable from speed by showing acceleration and speed as separate variables. This simulation features eliminate the cognitive difficulty of recognizing that acceleration is a derived measure (i.e., rate of change of speed) and helps students recognize that acceleration is directly affected by applied force as opposed to speed, therefore helping figure out that acceleration should be part of the mathematical relationship as opposed to speed. The PhET simulation also makes it easier to see the change in acceleration as a function of applied force rather than speed by clearly showing the numerical values for acceleration with varying applied force. Both aspects relate to recognizing unobserved variable of acceleration, which is usually convoluted with speed if phenomenon is observed in real life.

Blended sensemaking at the conceptual level with *Beer's Law* simulation involves recognizing that there is an additional variable that needs to be accounted for apart from concentration and container width to find the specific absorption properties of a substance. This variable is molar absorption coefficient (MAC), and it is not part of the PhET simulation on Beer's Law. Students can infer the information about this variable by noticing that different substances at a given wavelength and the same concentration absorb differently. Most students were provided MAC as part of the data on Beer's law simulation (see Appendix for data tables provided to students), and they could use the information on MAC to help them develop the exact mathematical relationship. At this level students justify the mathematical relationship they derived by using the numerical values of the relevant variables to show that they make sense for the specific form of the mathematical relationship.

Finally, *Energy Conversion* simulation blended sensemaking at the conceptual level involves recognizing the variable of "energy lost as thermal" in any system during the process of energy conversion. This thermal energy is not used for the purposes of the system (e.g., generating electricity). The PhET simulation shows thermal energy loss at every step of the process in the form of energy units leaving the system and not being used, but it is hard to notice this lost energy. This represents an unobserved variable for this phenomenon. The mathematical relationship can be derived either in the form of a) *Fraction of the Energy Used=Useful Energy output/Energy input* or b) *Useful Energy Output=Energy Input- Energy Lost to Useless Forms*. The mathematical relationship is justified by using the numerical values of the relevant variables to show that they make sense for the specific form of the mathematical relationship.

*Table 4. Sample responses for sub-level "description" for all levels of the framework.*

| Level | Acceleration | Beer's Law | Energy Conversion |
|-------|-------------|------------|-------------------|
| **1** **Qualitative** | "In the sim you can **change the mass and the force, and see how that affects acceleration**" | "I feel like the **variables that are really affecting absorbance are the concentration, the type of solution and the length**" | Interviewer: what are other important variables to include into that mathematical relationship? Student: **may be the starting amount of energy**. With the bicycle earlier, you make her paddle more, and you can see the heat leaving her. Then more and more energy gets put in, and then more energy ends up at the end heating the water. If you put less, it would be less at the end. Interviewer: any other energy element that is important to include? Student: **May be the final energy\***? The energy that is released after the function has been done. |
| **2** **Quantitative** | "**If you apply a force of 500 N and you have a mass of 50, then the net force is 500** (*no friction\*\**)**, you see that the acceleration is 10**" | "For absorbance, this one is a decimal number much smaller (*than transmittance*), the closer you are (*to the light source*), the smaller the number. **As you come up here** (*increase the distance to the light source*) **it comes from 0.16 to 1.3 (** *absorbance*)" | "Some of the energy is going away as thermal energy in the very beginning, but what energy IS going into the system into the generator is the same amount that's coming out. **So, if three things of mechanical energy go in, three things of mechanical energy come out**" |
| **3** **Conceptual** | "**The friction force was 84, in order to counter it, you would need, like 84 newtons of applied force**, and then the weight of this mass is 50 kg. **I solved on paper what the acceleration should be (**using F=ma**), and it should be 1.68 m/s², and that's what the sim is showing**" | Student: **I would guess there is something between the wavelength and the solution type, where there is, like, some constant, like a certain variable that is specific to the solution that determines what wavelength gets through**. Going back to the data, you would have to divide the absorbance and the length by the molar absorption coefficient in order to get the concentration. Interviewer: and why do you think you need to divide? Student: **just by manipulating the numbers, if I divided it every single time it would give me the correct number.** | Student: **I would say electrical energy equals ¼ thermal plus ⅜ of the mechanical.** Interviewer: can you explain why you derived it in that form? Student: because the input outlet has to be equal to the output. Interviewer: and if you were to generalize from this specific set-up to across set-ups, how would you change your equation? Student: **I think every energy will have a different way of showing stuff. The lightbulb would generate more heat, then, let's say, the fan. The ratio of the thermal energy to the ratio of the light energy would be a little different.** |

\* Bold text indicates key evidence in student responses for the specific sub-level
\*\*Italics text indicates clarification comments from the authors

Table 5 shows sample responses for each of the three disciplinary scenarios for the "Pattern" sub level at each of the broad levels of the framework. The "Pattern" sub-level at the lowest level, "qualitative", reflects student ability to identify the qualitative patterns among the observable variables relevant for characterizing the phenomenon. At this level students can't translate the identified qualitative patterns into the exact mathematical relationship. The next level, "quantitative", involves noticing the exact quantitative patterns among the observable variables. The quantitative patterns include recognizing direct and inverse relationships, or verbally describing a specific pattern (e.g., as variable A increases by X units, variable B increases by Y units). At this level, students can't translate the identified quantitative patterns into the exact mathematical relationship. Finally, at the highest level, "conceptual", students can translate the quantitative patterns they have noticed into the exact mathematical operations and develop a quantitative relationship for the phenomenon. They can fully explain the choice of the mathematical operations (and argue against choosing alternative mathematical operations using observations as evidence) and relate them to specific observations of the phenomenon. For example, sensemaking at the conceptual level for the *Acceleration* simulation would involve explicitly relating observations (larger mass requires more force to move) to the multiplication operation in the formula. For the *Beer's law* simulation, sensemaking at the conceptual level would involve explicitly relating the observations (increasing container width and concentration leads to increased absorbance) to the multiplication operation in the formula for absorption. Finally, for the *Energy Conversion* simulation, sensemaking at the conceptual level involves explicitly relating observations (useful energy is always a fraction of energy input) to the division (or multiplication operation if the conversion rate is known) in the formula. Students at this level can't provide a causal explanation for the equation structure (see table 6 conceptual level sample responses for comparison).

*Table 5. Sample responses for sub-level "patterns" for all levels of the framework.*

| Level | Acceleration | Beer's Law | Energy Conversion |
|---|---|---|---|
| **1**<br><br>**Qualit ative** | **"As you apply more force and the mass stays the same, that changes acceleration and makes it faster\*"** | "Transmittance and absorbance have sort of an opposite relationship, **as absorbance goes up transmittance goes down with increasing concentration**" | **"A lot more energy is being put into it than coming out"** |
| **2**<br><br>**Quant itative** | "The more mass that you get or the greater acceleration that you are going at, the greater the force will be. So, the force is **directly\*** related to the mass and the acceleration" | "I think the transmittance is **non-linear** with the concentration, and the absorbance is **linear** with the concentration when the wavelength is held at like a standard wavelength and the volume doesn't change (*container width\*\**)."<br>**"If it is 1 cm (*the vial width*), the maximum absorbance goes up by 0.5 every 100 mM"** | "It's not a 100% mechanical when it goes out because there is going to be the heat generated. **Every six electrical energy units that goes in, there is going to be one unit that comes out as a thermal** (*from the sim*)" |

| 3<br>Conceptual | "As the mass increases, it will require more force. **If you do like F=a/m** *(as opposed to F=ma)***, that means that as the mass increases, it will take less force to move the object**, which doesn't make sense, I think. As the mass increase, the object should be harder to move" | "As the concentration is higher, the absorbance rate gets higher as well, so that's proportional, on top. And then, opposite of the radius, as the radius gets...oh…. *(explores the sim)*...I was wrong in that. As the radius gets bigger, the absorbance will go up as well, **so it probably will be concentration times radius at a certain wavelength will equal to the rate of absorbance**" | Student: **energy output equals some sort of conversion rate, depending on what you are using, like a normal light bulb vs you are using like heating the water. The conversion rate represents the efficiency of the transfer of energy times, like, the input of the energy.** Interviewer: and you said that it is efficiency times the energy output. Why did you decide that it should be multiplication? Student: **in my mind, the efficiency was not a whole number, but a small number, like a percentage, so ranging from like a 1% to like a 100%. Multiplied just takes the maximum amount of energy that could be in a like ideas system, and then shrinks that to like what it actually is**. |

\* Bold text indicates key evidence in student responses for the specific sub-level
\*\*Italics text indicates clarification comments from the authors

Table 6 shows sample responses for each of the three disciplinary scenarios for the "Mechanism" sub level at each of the broad levels of the framework. As you can see, the "Mechanism" sub-level at the lowest level, qualitative, reflects student's ability to identify qualitative causal relationships among the relevant variables. For the *Acceleration* simulation this involves recognizing that applied force causes acceleration. For the *Beer's law* simulation, this would involve recognizing that concentration is the main causal factor behind changing absorbance. For the *Energy Conversion* simulation, this would involve recognizing that the reason all energy input cannot be converted into useful energy is because there is always thermal energy loss in the system. However, at this level students can't develop the exact mathematical relationship describing the phenomenon.

The "Mechanism" sub-level at the next level, "quantitative", reflects the student's ability to develop an exact quantitative relationship, justify that relationship using numerical values of the relevant variables, and recognize the qualitative causal mechanism behind the phenomenon. At this level students can't provide a causal explanation for the equation structure and can't justify the choice of the mathematical operation by directly relating their choice to patterns in their observations. That is what distinguishes it from the highest level, "conceptual". For the *Acceleration* simulation, this would involve developing the mathematical relationship for Newton's law, justifying the relationship using data (either from the simulation, or the data provided to the students), and recognizing that applied force causes acceleration. For the *Beer's law* simulation, this would involve developing a mathematical relationship for Beer's law, justifying the relationship using data (either from the sim, or the data provided to the students), and recognizing that change in concentration of the substance causes a change in the absorbance. Finally, for the *Energy Conversion* simulation, this would involve developing the mathematical relationship for efficiency in the form of *Fraction of Energy Used=Useful Energy output/Energy input* and justifying the equation using the data provided. Students at this level can qualitatively recognize that there is always energy lost as thermal in the process of energy conversion, but they struggle to relate it explicitly to the equation. In the example shown in table 6 the student recognized that the reason the LED lightbulb is more efficient than incandescent is because the incandescent lightbulb loses more thermal energy. However, when given wrong data that shows a system with over 100% efficiency (see the bottom right data table for *Energy Conversion* simulation in the Appendix) the student applied the formula derived earlier (*Efficiency=Useful Energy output/Useful Energy input*) and states that 125% efficiency is acceptable since it is a constant that holds across that system. This example demonstrates that while the student has qualitative understanding of the causal mechanism (recognizes that there is thermal energy lost from the system) and can derive the mathematical relationship for the phenomenon, they can't relate the mathematical relationship meaning to the causal mechanism. That is the distinguishing feature between this level and the highest level, conceptual.

      Finally, the "Mechanism" sub-level at the highest level, conceptual, reflects the ability to develop the exact quantitative relationship, justify the relationship by explicitly relating the choice of the mathematical operation to the observations, and provide causal explanation for the equation structure. For the *Acceleration* simulation, this would involve developing the mathematical relationship for Newton's law, justifying the choice of multiplication operation by directly relating observations (the force is directly related to mass and acceleration) to the choice of multiplication as the operation in the equation, and recognizing that the applied force causes acceleration. For the *Beer's law* simulation this would involve developing the mathematical relationship for Beer's law, justifying the choice of multiplication operation by directly relating observations (the absorbance is directly related to concentration and container width) to the choice of multiplication as the operation in the equation, and recognizing that concentration causes absorption. For *Energy Conversion*, this level would involve deriving the relationship for efficiency (either in the form of a) *Fraction of the Energy Used=Useful Energy output/Energy input* or b) *Useful Energy Output=Energy Input- Energy Lost to useless forms*) and relating the observations of energy lost to useless forms (such as thermal) to the formula derived across various systems (beyond specific cases).

*Table 6. Sample responses for sub-level "mechanism" for all levels of the framework.*

| Level | Acceleration | Beer's Law | Energy Conversion |
|---|---|---|---|
| **1**<br><br>**Qualit ative** | **"Acceleration would stop when you stop pushing, and you could see speed decrease"** | Interviewer: What is the main cause for what you observe? Student: **the amount of particles dissolved and the types of substance**. | **"I think everything has thermal in it, it is not all transferred into the kind of energy you actually want. There is going to be some thermal to it"** |
| **2**<br><br>**Quant itative** | Student: Let's see, in table A *(data table)* we did 250 divided by 50, we get five. **So, it's applied force minus friction over mass gives the acceleration**<br>Interviewer: Why did you state it in this form?<br>Student: **I guess because acceleration is the furthest right on the table** *(the data table)***, and usually in the equation what you try to figure out is on the right of the equal sign.**<br>Interviewer: what is the cause for all of these outcomes?<br>Student: **I think the applied force is the most important variable, it's the only thing that is really changing** | Student: I was able to figure out that the **absorbance is the molar absorption coefficient times the concentration times the container width**.<br>Interviewer: and why did you choose to express it that way?<br>Student: mostly because with a compound, the molar coefficient will be a constant, and **because the math worked out for the molar coefficient times the concentration time container width to equal the absorption**.<br>Interviewer: what do you think is the main causal factor behind changing absorbance and transmittance?<br>Student: **I'd say the concentration of the solution is the main changer.** | Student: I am assuming, the **LED, since it has more energy output as light, the incandescent light bulb is going to have more thermal energy than the LED.**<br>Interviewer: what is the percent efficiency now *(bottom right table)*?<br>Student: It's now a 125%.<br>Interviewer: Does that supports your model?<br>Student: yes, it still supports the model, it's just different.<br>Interviewer: why does it support the model?<br>Student: **because this percentage, in this case it's like 125%, the model took into account that there is a constant. So, like, it's a constant because it is 125% for this one, and 125% for this one** *(going down the column)*, **and so on and so forth. It took into account that there is a constant, it just the constant can change depending on the situation**. |
| **3**<br><br>**Conce ptual** | Student: **I think it's F=m*a because the more mass that you get or the greater acceleration that you are going at, the greater the force will be. So, the force is directly related to the mass and the acceleration.**<br>Interviewer: Do you think force is the outcome?<br>Student: Actually no, force is not the outcome. **The cause mechanism is a force which is causing the acceleration to go in a negative or positive direction, increase or decrease in magnitude**. | "**Maybe the absorbance has to do with multiplying by whatever the concentration is because multiplying by zero (when the concentration is zero) will give you zero absorbance**. Because no matter what all the other variables are, when concentration is zero absorbance is always zero. If it was addition or subtraction of the thickness of the container or the wavelength, it wouldn't matter what those are, if they get multiplied by concentration which is zero, you would still get zero. **If a concentration is zero, it's just water; adding anything to water makes it less clear; if anything is less clear, it makes it harder for the light to go through**" | "**In every equation there is going to be a different ratio of thermal energy in the right side of the equation.** Example- the light bulb (LED with solar panel). This one is very efficient. So, 1/8th of the total energy would be thermal, and the 7/8th that's produced will be light" |

\* Bold text indicates key evidence in student responses for the specific sub-level

\*\*Italics text indicates clarification comments from the authors

Table 7 shows final level assignment for all interviewed students for all three disciplinary scenarios. The general trend (15 students out of 25) was that students were assigned the same level and sub-level across the three disciplinary contexts. The other ten students exhibited different degree of variability in level assignment across the three disciplinary scenarios. Specifically, six students out of 25 were assigned the same broad level ("Qualitative", "Quantitative", "Conceptual") across all three disciplinary contexts (Physics, Chemistry, and Energy Conversion), but the sub-level assignment ("description", "pattern", "mechanism") varied within that level. The level assignments for these students are shown in italics in Table 7.

Further, the levels assignment for one of the students (student 2) differed by 1 sub-level only for one of the simulations. Specifically, student 2 was assigned level 1 "Mechanism" on *Acceleration* and *Beer's law* simulations but scored one sub-level higher at level 2 "Description" on the Energy Conversion simulation.

Only three students out of 25 were assigned different broad levels across the three disciplinary scenarios. The levels assignments for these students are shown in bold text in Table 7. Each had a unique notable feature. Student 3 was assigned level 3 Mechanism on *Acceleration* and *Energy Conversion* simulation but scored four sub-levels below at Level 2 Pattern on the *Beer's law* simulation. Notably, this student was an early interview and unlike the subsequent 22 of the others was not provided with the molar absorption coefficient on Beer's law simulation. We believe this affected his ability to bring together the quantitative observations made while interacting with the simulation to develop the exact mathematical relationship. Student 9 was assigned level 3 Mechanism on the *Acceleration* and *Beer's law* simulations but scored three sub-levels below at on *Energy Conversion* simulation. This student expressed a strong incoming pre-conception about the energy conversion process, which interfered with their sensemaking in the energy context. (The student was sure that an incandescent lightbulb uses thermal energy rather than electricity to produce light.) Finally, student 25 was assigned level 2 Mechanism on the *Acceleration* and *Energy Conversion* simulations but scored four sub-levels lower at level 1 Pattern on the *Beer's law* simulation. This student was very unfamiliar with the subject matter and seemed more confused than any of the other students as to what the simulation was showing. That appeared to affect their sensemaking during the exploration of this simulation.

In general, the data indicates that students tend to be assigned the same level and sub-level of the framework irrespective of disciplinary context. Most fluctuations happen for within level assignment where students score in different sub-levels (description, pattern, mechanism) of the same broad level (qualitative, quantitative, conceptual). Finally, it is rare that students are assigned sub-levels in different broad levels of the framework, but, as noted, this was usually because of some unique difficulty with one of the contexts Additionally, the level assignments seem to be reasonably well related with SAT Math scores below 650 but doesn't distinguish well for scores above 650 as shown in Table 7. We will discuss these findings and their implications in more detail.

*Table 7. Final level assignment for each simulation*

| Student | Acceleration | Beer's Law | Energy Conversion | SAT Math Score |
|---------|--------------|------------|-------------------|----------------|
| 1 | Level 3 Pattern | Level 3 Mechanism* | Level 3 Mechanism | Not available |
| 2 | <u>Level 1 Mechanism</u> | <u>Level 1 Mechanism*</u> | <u>Level 2 Description</u> | 540 |
| 3 | **Level 3 Mechanism** | **Level 2 Pattern*** | **Level 3 Mechanism** | 750 |
| 4 | *Level 2 Description* | *Level 2 Pattern* | *Level 2 Pattern* | 760 |
| 5 | Level 2 Pattern | Level 2 Pattern | Level 2 Pattern | 720 |
| 6 | *Level 3: Pattern* | *Level 3: Pattern* | *Level 3: Mechanism* | 680 |
| 7 | *Level 2: Mechanism* | *Level 2: Mechanism* | *Level 2: Pattern* | Not available |
| 8 | Level 2: Mechanism | Level 2: Mechanism | Level 2: Mechanism | 740 |
| 9 | **Level 3: Mechanism** | **Level 3: Mechanism** | **Level 2: Mechanism** | Not available |
| 10 | *Level 3: Pattern* | *Level 3: Mechanism* | *Level 3: Pattern* | Not taken |
| 11 | Level 3: Mechanism | Level 3: Mechanism | Level 3: Mechanism | 790 |
| 12 | *Level 3: Description* | *Level 3: Description* | *Level 3: Mechanism* | 560 |
| 13 | Level 3: Mechanism | Level 3: Mechanism | Level 3: Mechanism | 760 |
| 14 | Level 3: Mechanism | Level 3: Mechanism | Level 3: Mechanism | 730 |
| 15 | Level 3: Mechanism | Level 3: Mechanism | Level 3: Mechanism | 760 |
| 16 | Level 3: Mechanism | Level 3: Mechanism | Level 3: Mechanism | 690 |
| 17 | Level 3: Mechanism | Level 3: Mechanism | Level 3: Mechanism | 670 |
| 18 | Level 3: Mechanism | Level 3: Mechanism | Level 3: Mechanism | 660 |
| 19 | Level 3: Mechanism | Level 3: Mechanism | Level 3: Mechanism | 760 |
| 20 | Level 3: Mechanism | Level 3: Mechanism | Level 3: Mechanism | Not taken |
| 21 | Level 3: Mechanism | Level 3: Mechanism | Level 3: Mechanism | 710 |
| 22 | Level 1: Mechanism | Level 1: Mechanism | Level 1: Mechanism | 490 |
| 23 | *Level 1: Pattern* | *Level 1: Mechanism* | *Level 1: Mechanism* | 560 |
| 24 | Level 1: Pattern | Level 1: Pattern | Level 1: Pattern | 480 |
| 25 | **Level 2: Mechanism** | **Level 1: Pattern** | **Level 2: Mechanism** | 580 |

*Students were not provided molar absorption coefficient data

**Discussion**

In this paper we presented a theoretical framework for blended Math-Sci sensemaking grounded in prior research. The levels of the framework were developed following the blending process of the selected theoretical categories for Math and Science sensemaking dimensions originally described by Zhao and Schuchardt (2021). The final theoretical framework for blended Math-Sci sensemaking shown in Table 1 was reviewed by educational and subject matter experts and represents a cognition model reflecting qualitatively different ways of engaging in blended Math-Sci sensemaking process. The development of the framework helped answer the first RQ of our study: *How can one characterize the different ways of engaging in blended Math-Sci sensemaking?*

We gathered response process-based validity evidence for the theoretical framework by analyzing student responses from the interviews probing the levels of the theoretical framework shown in Table 1. The interviews were conducted in three disciplinary contexts, including Physics, Chemistry and Energy Conversion. The results of the interview analysis provided evidence for all the levels and sub-levels of the framework shown in Table 1. Specifically, we were able to identify evidence for all the different types of blended Math-Sci sensemaking in student responses for each disciplinary context, as illustrated in Table 4-6. This finding answers the second RQ of out study: *To what degree does the validity evidence support the theoretical framework for blended Math-Sci sensemaking?*

The framework shown in Table 1 represents a novel finding because it is the first detailed categorization of the blended Math-Sci sensemaking process that has been validated by student response data. This finding has important implications for instruction on blended Math-Sci sensemaking skills. Specifically, we have demonstrated that blended sensemaking can be characterized by different cognitive levels, which suggests that the framework shown in Table 1 can be used both as a diagnostic tool to accurately determine the cognitive of individual students as related to blended Math-Sci sensemaking, and as a guide for what needs to be emphasized during instruction to help students attain higher blended sensemaking ability (NRC, 2001).

Further, the data analysis showed that the level of student sensemaking tends to be consistent across the various disciplinary contexts, as shown in table 7. What little variation there is primarily occurs with sub-level assignment (description, pattern, mechanism) within a single broader level. The cross-level variation (qualitative, quantitative or conceptual) in the range of more than one sub-level occurs less often (total of 4 students out of 25) and occurred with cases that had certain distinguishing features such as lack of information provided to the other students (like student 3 who was not provided MAC), or strong preconceptions about the content area (student 9), or fundamental difficulty in understanding the phenomenon (student 25).

These findings suggest that blended Math-Sci sensemaking is a distinct cognitive construct irrespective of specific disciplinary context, which in turn has important implications for instruction. Specifically, it is likely that supporting students in developing Math-Sci sensemaking ability in one disciplinary context would help them apply such sense-making in other subjects. This hypothesis should be further investigated.

PhET simulations have features that make them uniquely effective research tools in this context. These simulations each provide an observable system governed by mathematical equations where the physical variables in those equations can be readily manipulated. Those manipulations produce changes in the behavior of the system that are immediately visible. In addition to providing visible qualitative changes in observables, many of the simulations also allow quantitative measures of the input and output quantities. This allows students to investigate qualitative, quantitative, and conceptual aspects of the phenomenon, therefore providing opportunities and encouragement to engage in blended sensemaking at all the levels described in the framework (Table 1). This is impossible to do with static representations of phenomena, such as text or drawings, and the use of real equipment to do this, while possible in principle, faces large practical and cognitive challenges. There are distracting complications in what is evident or not, and what can and cannot be easily manipulated.

The framework presented here will provide guidance for how to teach students to carry out blended Math-Sci sensemaking. Although it remains to be tested, it is likely that the levels of this framework serve as a learning progression for this type of sensemaking. Will students move efficiently from lower sensemaking levels to higher with appropriate learning experiences, and will they transfer this across different contexts? Exploring these questions will be the subject of future work.

The capabilities of PhET simulations that facilitated this research will likely also be useful for teaching sensemaking. This will be a focus of our future work. We will use the validated framework presented here as a guide to develop instructional sequences around PhET simulations focused on helping students develop blended Math-Sci sensemaking ability in various scientific disciplines. We will create instructional sequences with practice tasks and questions to answer that follow the framework progression.

Another area of future work is to extend this work to create an assessment instrument to easily and accurately diagnose individual student's level of blended Math-Sci sensemaking. We will specifically align individual assessment items to the sub-levels of the framework to probe student blended sensemaking ability at each individual sub-level. We hope that the diagnostic tool aligned with the instructional sequences through the framework presented here will be used to support students in developing higher proficiency in blended sensemaking.

## Acknowledgements
We would like to thank Dr. Joselyn Nardo for helping conduct inter-rater reliability analysis and the PhET project for making this work possible by providing open-source simulations.

## Disclosure statement
The authors have no conflict of interest to disclose

## Funding
This work is funded by the Yidan foundation

# Appendix

A1. Interview Protocol

**<u>General Question:</u>**
1. Explore the simulations:
    i. Physics: "Acceleration" sim.
    ii. Chemistry: "Beer's law" sim.
    iii. Climate Science: "Systems" sim.
    iv. Baseline: "Torque" sim.
Give the student some time to explore all simulations
2. What did you observe in the simulations?
3. What is the phenomenon described in the sim?
4. Your task is to describe the "Acceleration"/ "Beer's law"/ "Systems"/"Torque" sim using a mathematical formula. How would you go about it?

**<u>Level 1 (Qualitative):</u>**
5. What are the variables and what are the constants relevant for characterizing the phenomenon mathematically?
6. How does each variable relate to what is happening in the sim?
7. What is the qualitative relationship between the variables you identified?

**<u>Level 2 (Quantitative):</u>**
8. What is the quantitative relationship?
    a. Can you describe the relationship in terms of a specific mathematical relationship?
    b. How would you go about establishing the quantitative relationship?
    c. What are important variables to measure?
    d. Why did you choose to express the relationship in that form? Why is it not "+";"-"; "/" or " * "?
9. <u>Share the data related to the simulation with the student.</u> Explore the data provided to you (*see data given to students for each sim on next page*).
    a. Based on the data, what is your suggested quantitative relationship between the variables?
    b. How would you interpret this data in terms of quantitative relationship?
    c. Is there any information that you are missing to determine the exact quantitative relationship between the variables? d. Why did you choose to relate the variables that way?
10. <u>If student still struggles to derive the formula</u>, share a list of possible formulas with them (*see next page for formulas given)*
    a. Which formula is the most likely and why?

**<u>Level 3 (Conceptual):</u>**
11. How would you describe the causal mechanism of the phenomenon in the sim using both your interactions with the sim and the mathematical formula you have developed? If unclear, explain the meaning of "causal".
12. How does the mathematical operation you chose for your equation relate to the observations?
13. What is/are the important observable outcome(s) of the phenomenon?
14. What is the cause of the observed outcome(s)?
15. How do you quantitatively explain the cause of the observed outcome?
16. Have you identified any extreme cases? If yes, how would you explain what causes the extreme cases using formula?

# Acceleration Simulation Data and Formulas

## Data

Physics: which data supports your model (Data from Sim)? You can use calculator, paper and pencil if needed

**Table A**

| Mass (kg) | Applied Force (N) | Friction (N) | Acceleration (m/s^2) |
|---|---|---|---|
| 50 | 150 | 94 | 1.12 |
| 50 | 200 | 94 | 2.12 |
| 50 | 250 | 94 | 3.12 |
| 50 | 300 | 94 | 4.12 |

**Table C**

| Object | Mass (kg) | Applied Force (N) | Friction (N) | Acceleration (m/s^2) |
|---|---|---|---|---|
| Fridge + 2 wooden boxes | 300 | 188 | 88 | 0.33 |
| Fridge + water bucket | 300 | 188 | 88 | 0.33 |
| Fridge | 200 | 186 | 86 | 0.5 |
| Water bucket + 2 wooden boxes | 200 | 186 | 86 | 0.5 |

**Table B**

| Mass (kg) | Applied Force (N) | Friction (N) | Acceleration (m/s^2) |
|---|---|---|---|
| 50 | 150 | 0 | 3 |
| 50 | 200 | 0 | 4 |
| 50 | 250 | 0 | 5 |
| 50 | 300 | 0 | 6 |

**Possible Formulas**

Which formula/formulas best support your observations in the simulation?

A)  F = a/m
B)  F = m + a
C)  F = m/a
D)  F= m - a
E)  F= m * a

F) a=F/m
G) a=m/F
H) a=F+m
I) a=F-m
J) a= F * m

**Beer's Law Simulation Data and Formulas**

## Data

Does that data support your model ?

Compound A at 549 nm

| Conc. (mol/L) | A | T (%) | l (cm) | Molar Absorption Coefficient | Temperature (C) |
|---|---|---|---|---|---|
| 0.01 | 0.03 | 93.52 | 1 | 3 | 25 |
| 0.05 | 0.15 | 71.34 | 1 | 3 | 25 |
| 0.1 | 0.3 | 51.31 | 1 | 3 | 25 |
| 0.15 | 0.45 | 36.65 | 1 | 3 | 25 |
| 0.2 | 0.6 | 26.23 | 1 | 3 | 25 |
| 0.25 | 0.75 | 18.77 | 1 | 3 | 25 |

Compound B at 549 nm

| Conc. (mol/L) | A | T (%) | l (cm) | Molar Absorption Coefficient | Temperature (C) |
|---|---|---|---|---|---|
| 0.01 | 0.04 | 90.25 | 1 | 4 | 25 |
| 0.05 | 0.2 | 60.15 | 1 | 4 | 25 |
| 0.1 | 0.4 | 35.82 | 1 | 4 | 25 |
| 0.15 | 0.6 | 21.57 | 1 | 4 | 25 |
| 0.2 | 0.8 | 13.46 | 1 | 4 | 25 |
| 0.25 | 1 | 8.91 | 1 | 4 | 25 |

Which relationship (Abs vs Conc or Transmit. vs Conc) is easier to use to determine concentration?

## Possible Formulas

Which formula/formulas best support your observations in the simulation?

a) A= Concentration * Transmittance
b) A= Concentration/Transmittance
c) A=Concentration* Transmittance * Vial Width
d) A=Transmittance + Vial Width
e) A= Transmittance/Vial Width
f) A= Transmittance * Vial Width * Molar Absorption Coefficient
g) A= Concentration * Vial Width * Molar Absorption Coefficient
h) A = Vial Width/Molar Absorption Coefficient
i) A= Concentration * Vial Width * Molar Absorption Coefficient * Transmittance
j) A= Transmittance - (Concentration * Vial Width* Molar Absorbance Coefficient)

# Energy Conversion Simulation Data and Formulas

## Data

### Incandescent light Bulb

| Calories consumed (J) | Energy Generated by the generator wheel (J) | Bulb Energy Output (J) | Life (Years) |
|---|---|---|---|
| 100 | 50 | 12.5 | 0.9 |
| 200 | 100 | 25 | 0.9 |
| 300 | 150 | 37.5 | 0.9 |
| 400 | 200 | 50 | 0.9 |
| 500 | 250 | 62.5 | 0.9 |

### LED light Bulb

| Calories consumed (J) | Energy Input in generator wheel (J) | Energy Output (J) | Life (Years) |
|---|---|---|---|
| 100 | 50 | 25 | 22.8 |
| 200 | 100 | 50 | 22.8 |
| 300 | 150 | 75 | 22.8 |
| 400 | 200 | 100 | 22.8 |
| 500 | 250 | 125 | 22.8 |

| Light Hitting the Solar Panel (J) | Energy Generated by the Solar Panel (J) | Bulb Energy Output (J) | Life (Years) |
|---|---|---|---|
| 100 | 20 | 12.5 | 0.9 |
| 200 | 40 | 25 | 0.9 |
| 300 | 60 | 37.5 | 0.9 |
| 400 | 80 | 50 | 0.9 |
| 500 | 100 | 62.5 | 0.9 |

| Light Hitting the Solar Panel (J) | Energy Generated by the Solar Panel (J) | Bulb Energy Output (J) | Life (Years) |
|---|---|---|---|
| 100 | 20 | 25 | 22.8 |
| 200 | 40 | 50 | 22.8 |
| 300 | 60 | 75 | 22.8 |
| 400 | 80 | 100 | 22.8 |
| 500 | 100 | 125 | 22.8 |

## Possible Formulas

Which formula/formulas best support your observations in the simulation?

a) Useful Energy = Energy Output * Energy Input

b) Useful Energy = Energy Output + Energy Input

c) Useful Energy = Energy Input / Energy Output

d) Useful Energy = Energy Output/Energy Input

e) Useful Energy % = (Energy Output/Energy Input) * 100%

f) Useful Energy = Energy Input - Energy Output

g) Useful Energy = Energy Output - Energy Input

h) Useful Energy % = (Energy Output/Energy Input) / 100%

i) Useful Energy % = (Energy Output/Energy Input) + 100%

j) Useful Energy % = (Energy Output/Energy Input) - 100%

*A2. Interview Coding Rubric for Acceleration Simulation.*

| 1 Qualitative | Description | Students identify that applied force and mass affect acceleration and speed. |
|---|---|---|
| | Pattern | Student identifies that for constant mass, larger applied force results in larger speed and acceleration; for constant force, the larger the mass, the smaller the acceleration and speed. |
| | Mechanism | Students recognize that applied force causes acceleration of an object in the same direction but can't define the exact mathematical relationship for acceleration as a function of applied force. |
| 2 Quantitative | Description | Students recognize that mass and applied force affect acceleration; they describe the relationship quantitatively by reiterating the data from the sim (when I use mass X, acceleration changes to Y; when I apply force X, acceleration changes to Y etc.) |
| | Pattern | Students recognize that the relationship is positive linear between applied force and acceleration and applied force and mass. Student explains that if they increase the applied force by X units acceleration increases linearly by Y units. Students not yet able to relate the observed patterns to the operations in a mathematical equation and can't develop exact mathematical relationship for Newton's law yet. |
| | Mechanism | Students develop the exact mathematical relationship for acceleration either in the form of $a=F_{net}/m$ or $F_{net}=m*a$ (where $F_{net}=F_{applied}-F_{friction}$) and justify the formula using numerical values for the variables obtained either from the simulation or from the data provided to them. Students struggle to completely explain why the mathematical formula makes sense (explain why multiplication or division make sense and why subtraction or addition doesn't make sense in the formula). *Students might also struggle to explain why they chose to express the formula using acceleration as opposed to speed, and why $F_{net}$ is different from $F_{applied}$ if not using PhET simulation.*<br><br>*Note: PhET simulation shows acceleration and speed as separate variables, therefore reducing the cognitive difficulty of recognizing that acceleration is a derived measure. The simulation also makes it easier to see the change in acceleration as a function of applied force rather than speed. These aspects relate to recognizing unobserved variable of acceleration, which is usually convoluted with speed if phenomenon is observed in real life. Additionally, the simulation provides an easy way to observe applied force and total force ($F_{applied}-F_{friction}$) as separate variables, therefore reducing the difficulty of having to recognize an additional variable ($F_{friction}$) that should be a part of the equation. These simulation design features simplify the task of recognizing* |
| | | *unobserved variables of acceleration and $F_{friction}$.* |
| 3 Conceptual | Description | Students recognize that acceleration is a derived measure and can develop the exact mathematical relationship either in the form of a=$F_{net}$/m or $F_{net}$=m*a. Students justify the multiplication operation in the formula using numerical values for the variables obtained either from the simulation or from the data provided to them. Students recognize that $F_{net}$ is calculated by subtracting applied force from the force of friction and recognize that friction force relates to specific surface qualities. |
| | Pattern | Students recognize that acceleration is a derived measure and can develop the exact mathematical relationship either in the form of a=$F_{net}$/m or $F_{net}$=m*a. Students justify the multiplication operation in the formula by recognizing that the relationship is positive linear between applied force and acceleration and applied force and mass. Similarly, students can explain why F= m/a or F= m+a or F=m-a DOES NOT make sense by connecting their math understanding to the simulation. |
| | Mechanism | Students recognize that the *net force distributed over mass* causes the acceleration of an object in the same direction; recognize that acceleration is a derived measure and therefore can develop the exact mathematical relationship either in the form of a=$F_{net}$/m or $F_{net}$=ma as well as explain why multiplication or division make sense and why subtraction or addition doesn't make sense in the formula. Students can explain why they chose acceleration rather than speed to be a variable in the equation and can explain why $F_{net}$=$F_{applied}$-$F_{friction}$. |

*A3. Interview Coding Rubric for Energy Forms Simulation.*

| 1 Qualitative | Description | Students identify energy input and energy output as the key variables. |
|---|---|---|
| | Pattern | Students identify that larger energy input results in larger energy output and that energy output is always smaller than energy input. |
| | Mechanism | Students recognize that all the consumed energy never turns into useful energy. |
| 2 Quantitative | Description | Students recognize that the useful energy is always less than energy input; they describe the relationship quantitatively by reiterating the data from the sim (when X energy units (squares in the sim) go into a specific energy generator, Y energy units get converted into useful energy and Z energy units get lost to the surroundings) |
| | Pattern | Students recognize that useful energy is always smaller than energy input; they recognize that for every certain number of useful energy units generated, there is always a certain amount of energy lost as heat; they can't develop the exact general mathematical relationship that applied across all systems yet; they might call this efficiency. |
| | Mechanism | Students can explain *quantitatively* (derive exact mathematical relationship that applied across all systems) for how the energy input relates to energy output (output energy/energy input=fraction of energy used) and justify the formula by using the simulation or the data provided to them. *Students struggle to fully explain what the formula means and relate to the idea of inevitable energy loss- difference between this level and Concept. Mechanism.* |
| 3 Concep | Description | Students can derive the formula for efficiency in the form: useful energy/energy input=fraction of energy used OR energy input=useful energy-energy lost. Students recognize that in all processes there is a fraction of energy lost to the surroundings and not converted to useful energy. Students justify the multiplication operation in the formula using numerical values for the variables obtained either from the simulation or from the data provided. <u>*Note*</u>: *PhET simulation shows thermal energy loss at every step of the system qualitatively in the form of energy units leaving the system and not being used at the end, but it is hard to notice right away. This variable represents an unobserved variable for this phenomenon.* |
| | Pattern | Students can develop the exact mathematical relationship for efficiency either in the form useful energy/energy |

| t u al | | input=fraction of energy used OR energy input=useful energy-energy lost. Students recognize that in all processes there is a fraction of energy lost to the surroundings and not converted to useful energy. Students can explain the division operation in the formula by recognizing that useful energy is always a fraction of energy input, which suggests division. |
|---|---|---|
| | Mechanism | Students can develop the exact mathematical relationship for efficiency that applied across systems and explain the patterns among variables in the formula (useful energy is always a fraction of energy input). Students recognize that useful energy is always a fraction of energy input because some energy is always lost to the surroundings (for example, in the form of thermal energy). Students recognize that not all systems are equally efficient and recognize that energy loss will be different for different systems. Students recognize than energy efficiency can never be 100% of above. |